\documentclass[prl,aps,twocolumn]{revtex4}

\usepackage{psfrag}
\usepackage{graphicx}
\usepackage{color}
\usepackage{dcolumn}
\usepackage{latexsym,amsfonts}
\usepackage{bm}
\usepackage{amssymb}
\begin{document}

\title{Time Modulation of Orbital Electron Capture Decays of H--like
Heavy Ions}

\author{A. N. Ivanov${^a}$, P. Kienle$^{b,c}$}
\affiliation{${^a}$Atominstitut der \"Osterreichischen
Universit\"aten, Technische Universit\"at Wien, Wiedner Hauptstrasse
8-10, A-1040 Wien, Austria} \affiliation{${^b}$Stefan Meyer Institut
f\"ur subatomare Physik \"Osterreichische Akademie der Wissenschaften,
Boltzmanngasse 3, A-1090, Wien, Austria} \affiliation{${^c}$Excellence
Cluster Universe Technische Universit\"at M\"unchen, D-85748 Garching,
Germany} \email{ivanov@kph.tuwien.ac.at}

\date{\today}

\begin{abstract}
  According to experimental data at GSI, the rates of the number of
  daughter ions, produced by the nuclear K--shell electron capture
  ($EC$) decays of the H--like ${^{140}}{\rm Pr}^{58+}$, ${^{142}}{\rm
  Pm}^{60+}$ and ${^{122}}{\rm I}^{52+}$ ions, are modulated in time
  with periods $T_{EC}$ of the order of a few seconds, obeying an
  $A$--scaling $T_{EC} = A/20\,{\rm s}$, where $A$ is the mass number
  of the mother nuclei, and with amplitudes $a^{EC}_d \sim 0.21$. In
  turn, the positron decay mode of the H--like ${^{142}}{\rm
  Pm}^{60+}$ ions showed no time modulation of the decay rates. As has
  been shown in Phys. Rev. Lett. {\bf 103}, 062502 (2009) and
  Phys. Rev. Lett. {\bf 101}, 182501 (2008), these data can be
  explained by the interference of two massive neutrino
  mass--eigenstates. In this letter we give a reply on the comments on
  our paper Phys. Rev. Lett. {\bf 103}, 062502 (2009) by A. Gal
  (arXiv: 0809.1213v2). PACS: 12.15.Ff, 13.15.+g, 23.40.Bw, 26.65.+t
\end{abstract}

\maketitle

\subsubsection*{``GSI Oscillations'' as interference of massive neutrino 
mass--eigenstates \cite{Ivanov2}}

The experimental data on the K--shell electron capture ($EC$) and
 positron ($\beta^+$) decay rates of the H--like heavy ions, carried
 out in the Experimental Storage Ring (ESR) at GSI in Darmstadt
 \cite{GSI2}--\cite{GSI5}, place the following constraints on the
 theoretical approaches for explaining of these phenomena: 1) the
 periods $T_{EC}$ of the time modulation of the $EC$--decay rates obey
 the $A$--scaling $T_{EC} = A/20\,{\rm s}$, where $A$ is the mass
 number of the mother ion, 2) the modulation amplitude for all
 observed decays is $a^{EC}_d \simeq 0.21$ and 3) the $\beta^+$--decay
 rates have no time modulated terms. In addition all theoretical
 approaches, describing the experimental data of the $EC$ and
 $\beta^+$ decay rates of the H--like heavy ions
 \cite{GSI2}--\cite{GSI5}, should explain the absence of the time
 modulation of the $EC$--decay rates of bound atoms ${^{142}}{\rm Pm}$ and
 ${^{180}}{\rm Re}$, measured in \cite{Pm,Re}.

As has been shown in \cite{Ivanov2}--\cite{Ivanov7}, the experimental
data \cite{GSI2}--\cite{GSI5}, called the ``GSI Oscillations'', and
the observations reported in \cite{Pm,Re} can be explained only
following the hypothesis of the interference of the massive neutrino
mass--eigenstates $|\nu_j\rangle$ \cite{Ivanov2,Ivanov4}, defining the
neutrino $|\nu_{\alpha}\rangle$ with a lepton charge $\alpha = e, \mu$
or $\tau$ as a coherent superposition of massive neutrino
mass--eigenstates $|\nu_{\alpha}\rangle = \sum_jU^*_{\alpha
j}|\nu_j\rangle$ \cite{PDG08}. Neither the magnetic field of the ESR
\cite{Ivanov6} (see also \cite{Faestermann}) nor the mass--splitting
of the H--like mother ions \cite{Ivanov5,Ivanov7} (see also
\cite{Ivanov2}) enable to explain the experimental data
\cite{GSI2}--\cite{Re}. The discussions of the ``GSI oscillations'' as
the interference of massive neutrino mass--eigenstates were proposed
in \cite{GSI}. 

Recently \cite{Gal2}, Gal has criticised the results obtained in
\cite{Ivanov2}. The main aim of this letter is to clarify all
problems, which have prevented Gal from accepting our results.

\subsubsection*{Frequencies of ``GSI oscillations'', 
caused by interferences of massive neutrino mass--eigenstates, should
be inversely proportional to neutrino energies, i.e. to $Q$--values of
$EC$--decays of H--like heavy ions, but not to the mass number $A$ of
mother ions \cite{Gal2}}

The amplitude of the $EC$--decay $m \to d + \nu_e$, caused by a
Gamow--Teller $1^+ \to 0^+$ transition of the mother ion $m$ from the
ground hyperfine state $(1s)_{F = \frac{1}{2}}$ into the daughter ion
in the stable ground state and the electron neutrino $\nu_e$, a
coherent superposition $|\nu_e\rangle = \sum_jU^*_{e j}|\nu_j\rangle$
of massive neutrino mass--eigenstates $\nu_j$ with masses $m_j$, is a
function of time $t$ defined by \cite{Ivanov2}
\begin{eqnarray}\label{label1}
 A(m \to d \,\nu_e)(t) = \sum_jU_{e j}A(m \to d \,\nu_j)(t),
\end{eqnarray}
where the amplitude $A(m \to d \,\nu_j)(t)$ of the $m \to d + \nu_j$
transition is calculated with the Hamilton operator of weak
interactions given by \cite{Ivanov2,Ivanov4}
\begin{eqnarray}\label{label2}
\hspace{-0.3in}&&{\rm H}^{(j)}_W(t)
  =\frac{G_F}{\sqrt{2}}V_{ud}\!\!\int\!\!  d^3x
  [\bar{\psi}_n(x)\gamma^{\mu}(1 - g_A\gamma^5) \psi_p(x)]\nonumber\\
\hspace{-0.3in}&& \times [\bar{\psi}_{\nu_j}(x)
  \gamma_{\mu}(1 - \gamma^5)\psi_{e^-}(x)],
\end{eqnarray}
with standard notation \cite{Ivanov2,Ivanov4,Ivanov1}. It is equal to
\begin{eqnarray}\label{label3}
  \hspace{-0.3in}&&A(m \to d\, \nu_j)(t) = -
  \,\delta_{M_F,-\frac{1}{2}}\,\sqrt{3} \sqrt{2 M_m } \nonumber\\
\hspace{-0.3in}&&\times\,{\cal M}_{\rm GT}\,\langle
  \psi^{(Z)}_{1s}\rangle\, \sqrt{2 E_d(\vec{q}_j)
  E_j(\vec{k}_j)}\,\frac{e^{\,i(\Delta E_j - i\varepsilon)t}}{\Delta
  E_j - i\varepsilon}\nonumber\\
   \hspace{-0.3in}&&\times \,\Phi_d(\vec{k}_j + \vec{q}_j),
\end{eqnarray}
where $\Delta E_j = E_d(\vec{q}_j) + E_j(\vec{k}_j) - M_m$ is the
energy difference of the final and initial state, $M_m$ is the mother
ion mass, $E_d(\vec{q}_j)$ and $E_j(\vec{k}_j)$ are the energies of
the daughter ion and massive neutrino $\nu_j$ with 3--momenta
$\vec{q}_j$ and $\vec{k}_j$, respectively, ${\cal M}_{\rm GT}$ is the
nuclear matrix element of the Gamow--Teller transition $m\to d$ and
$\langle \psi^{(Z)}_{1s}\rangle$ is the wave function of the bound
electron in the H--like heavy ion $m$, averaged over the nuclear
density \cite{Ivanov1}.

Suppose that in the $EC$--decay $m \to d + \nu_e$ the momenta of the
 daughter ions $\vec{q}_j$, produced in the decay channels $m \to d +
 \nu_j$, are precisely measured. In case that the differences of
 momenta $|\vec{q}_i - \vec{q}_j|$ are larger then the momentum
 resolutions $|\delta \vec{q}_j|$, i.e.  $|\vec{q}_i - \vec{q}_j| \gg
 |\delta \vec{q}_j|$, where $\vec{q}_i$ is a 3--momentum of a daughter
 ion in the decay channel $m \to d + \nu_i$ for $i\neq j$, all decay
 channels $m \to d + \nu_j$ are distinguished experimentally and the
 $EC$--decay rate should never show time modulation. This agrees with
 the assertion pointed out in Ref.\cite{Glashow}.

However, this is not the case with the ``GSI oscillations''
\cite{GSI2}--\cite{GSI5}. The wave function of the detected daughter
ion should be taken in the form of a wave packet, since the time
differential detection of the daughter ions from the $EC$--decays with
a time resolution $\tau_d \simeq 0.32\,{\rm s}$ introduces energy
$\delta E_d \sim 2\pi/\tau_d$ and 3--momentum $|\delta \vec{q}_d| \sim
2\pi/\tau_dv_d$ uncertainties, where $v_d$ is the velocity of the
daughter ion in the ESR \cite{Ivanov2}. Such a smearing is described
by the wave function $\Phi_d(\vec{k}_j + \vec{q}_j)$ \cite{Ivanov2}.
In the rest frame of the H--like mother ion the energy and momentum
uncertainties are equal to $\delta E_d \sim 2\pi\gamma /\tau_d =
1.85\times 10^{-14}\,{\rm eV}$ and $|\delta \vec{q}_d| \sim 2\pi
\gamma M_d/\tau_d Q_{EC}$, where $\gamma = 1.432$ is the Lorentz
factor of the H--like mother ions \cite{GSI2}, $v_d = Q_{EC}/M_d$ is a
velocity of the daughter ion and $Q_{EC}$ is the $Q$--value of the
$EC$--decay, equal to the momentum of the daughter, and $M_d$ is the
mass of the daughter ion. For the $EC$--decay ${^{140}}{\rm Pr}^{58+}
\to {^{140}}{\rm Ce}^{58+} + \nu_e$ the $Q$--value is equal to $Q_{EC}
= 3348(6)\,{\rm keV}$ \cite{Ivanov1}. This gives $|\delta \vec{q}_d|
\sim 7.21\times 10^{-10}\,{\rm eV}$.

Due to energy and momentum conservation in every $EC$--decay channel
$m \to d + \nu_j$ the energy and momentum of massive neutrino $\nu_j$
are  equal to \cite{Ivanov2}
\begin{eqnarray}\label{label4}
E_j(\vec{k}_j) \simeq Q_{EC} + \frac{m^2_j}{2
  M_m}\,,\, |\vec{k}_j|\simeq Q_{EC} - \frac{m^2_j}{2 Q_{EC}},
\end{eqnarray}
where $Q_{EC} = M_m - M_d$. The differences of energies and momenta of
neutrino mass--eigenstates are
\begin{eqnarray}\label{label5}
\omega_{ij} &=& E_i(\vec{k}_i) - E_j(\vec{k}_j) = \frac{\Delta
m^2_{ij}}{2 M_m},\nonumber\\ k_{ij} &=& |\vec{k}_i| - |\vec{k}_j| =
-\,\frac{\Delta m^2_{ij}}{2 Q_{EC}}.
\end{eqnarray}
In turn, $\omega_{ij}$ and $k_{ij}$ determine also the recoil energy
and 3--momentum differences of the  daughter ions.

For two massive neutrino mass--eigenstates and the $EC$--decay of
${^{140}}{\rm Pr}^{58+}$ \cite{Ivanov2} we get
\begin{eqnarray}\label{label6}
\omega_{21} &=&\frac{\Delta m^2_{21}}{2 M_m} = 8.40\times
10^{-16}\,{\rm eV},\nonumber\\ |k_{21}| &=& \frac{\Delta m^2_{21}}{2
Q_{EC}} = 3.27\times 10^{-11}\,{\rm eV},
\end{eqnarray}
where we have set $\Delta m^2_{21} = 2.19\times 10^{-4}\,{\rm eV^2}$
\cite{Ivanov2}. Since $\delta E_d \gg \omega_{21}$ and $|\delta
\vec{q}_d| \gg |k_{21}|$, the daughter ions, produced in the two decay
channels $m \to d + \nu_1$ and $m \to d + \nu_2$, are
indistinguishable \cite{Ivanov2}. This is the origin of the coherence
in the $EC$--decays $m \to d + \nu_e$ of the H--like heavy ions,
measured in GSI \cite{Ivanov2}.

Thus, in GSI experiments the observed daughter ion $d$ is a nucleus
with energy $E_d(\vec{q}\,)$ and 3--momentum $\vec{q}$ for all decay
channels $m \to d + \nu_j$.  The amplitude of the $m \to d + \nu_e$
decay reads
\begin{eqnarray}\label{label7}
  \hspace{-0.3in}&&A(m \to d \, \nu_e)(t) = -
  \delta_{M_F,-\frac{1}{2}}\,\sqrt{3} \sqrt{2 M_m}{\cal M}_{\rm GT}
  \nonumber\\
\hspace{-0.3in}&&\times\,\langle
  \psi^{(Z)}_{1s}\rangle \sum_j U_{ej} \sqrt{2 E_d(\vec{q}\,)
  E_j(\vec{k}_j)}\,\frac{e^{\,i(\Delta E'_j -
  i\varepsilon)t}}{\Delta E'_j -
  i\varepsilon}\nonumber\\
   \hspace{-0.3in}&&\times \,\Phi_d(\vec{k}_j + \vec{q}\,),
\end{eqnarray}
where $\Delta E'_j = E_d(\vec{q}\,) + E_j(\vec{k}_j) - M_m$. The
$EC$--decay rate is related to the expression
\begin{eqnarray}\label{label8}
\hspace{-0.3in}&&\lim_{\varepsilon \to
0}\frac{d}{dt}\frac{1}{2}\sum_{M_F}|A(m\to d \, \nu_e)(t)|^2 = 3 M_m
|{\cal M}_{\rm GT}|^2 \nonumber\\
\hspace{-0.3in}&&\times\,|\langle \psi^{(Z)}_{1s}\rangle|^2 \Big\{\sum_{j =
1,2}|U_{ej}|^2 2 E_d(\vec{q}\,)E_j(\vec{k}_j)\,2\pi\,\delta(\Delta
E'_j)\nonumber\\
\hspace{-0.3in}&&\times |\Phi_d(\vec{k}_j + \vec{q}\,)|^2 + \sum_{i >
j}U^*_{e i}U_{ej}\sqrt{ 2 E_d(\vec{q}\,)E_i(\vec{k}_i)}\nonumber\\
\hspace{-0.3in}&&\times \sqrt{2
E_d(\vec{q}\,)E_j(\vec{k}_j)}\,\Phi^*_d(\vec{k}_i +
\vec{q}\,)\,\Phi_d(\vec{k}_j + \vec{q}\,)\nonumber\\
\hspace{-0.3in}&&\times\,[2\pi\,\delta(\Delta E'_i) +
2\pi\,\delta(\Delta E'_j)]\,\cos(\omega_{ij}t)\Big\}.
\end{eqnarray}
This expression reproduces Eq.(\ref{label9}) in our paper
\cite{Ivanov2} with the same frequencies $\omega_{ij} = \Delta
m^2_{ij}/2 M_m$.  

As has been mentioned in \cite{Ivanov2}, the first term in
Eq.(\ref{label8}) is the sum of the two diagonal terms of the
transition probability into the states $d + \nu_1$ and $d + \nu_2$,
describing the incoherent contribution of massive neutrino
mass--eigenstates, while the second term defines the interference of
states $ \nu_i$ and $ \nu_j$ with $i\neq j$ causing the periodic time
dependence with the frequency $\omega_{ij}$ equal to
\begin{eqnarray}\label{label9} 
&&\omega_{ij} = \Delta E'_i - \Delta E'_j = E_d(\vec{q}\,) +
E_i(\vec{k}_i) - M_m \nonumber\\ &&- E_d(\vec{q}\,) - E_j(\vec{k}_j) +
M_m = E_i(\vec{k}_i) - E_j(\vec{k}_j) =\nonumber\\ &&= \frac{\Delta
m^2_{ij}}{2 M_m.}
\end{eqnarray}
Thus, we argue that the interference term, produced by a coherent
contribution of massive neutrino mass--eigenstates with a frequency
inversely proportional to the mass number $A$ of the mother ion, can
be observed only due to energy and momentum uncertainties, introduced
by the time differential detection of the daughter ions
\cite{Ivanov2}.

Due to the smallness of neutrino masses for the calculation of the
$EC$--decay rate we can take the massless limit everywhere except the
modulated term $U^*_{e i}U_{ej}\cos(\omega_{ij}t)$
\cite{Ivanov2}. Since the 3--momenta of the massive neutrino
mass--eigenstates and the 3--momenta of the daughter ions differ only
slightly from the $Q$--value of the $EC$--decay, we can set $\vec{k}_i
\simeq \vec{k}_j \simeq \vec{k}$ and $|\Phi_d(\vec{k} + \vec{q}\,)|^2
= V (2\pi)^3\,\delta^{(3)}(\vec{k} + \vec{q}\,)$, where $V$ is a
normalisation volume \cite{Ivanov2}. Using the definition of the
$EC$--decay rate
\begin{eqnarray}\label{label10}
\hspace{-0.3in}&& \lambda_{EC}(t) = \frac{1}{2M_m V}\int
\frac{d^3q}{(2\pi)^3 2E_d}\frac{d^3k}{(2\pi)^3
2E_{\nu_e}}\nonumber\\
\hspace{-0.3in}&& \times\,\lim_{\varepsilon \to
0}\frac{d}{dt}\frac{1}{2}\sum_{M_F}|A(m\to
d\,\nu_e)(t)|^2,
\end{eqnarray}
we obtain the following expression for the time modulated $EC$--decay
rate \cite{Ivanov2}
\begin{eqnarray}\label{label11}
\lambda_{EC}(t) = \lambda_{EC}(1 +
a_{EC}\,\cos(\omega_{21}t)),
\end{eqnarray}
where $\omega_{21} = \Delta m^2_{21}/2M_m$, $a_{EC} =
\sin2\theta_{12}$ and $\lambda_{EC}$ has been calculated in
\cite{Ivanov1}. The $EC$--decay rate Eq.(\ref{label11}) is calculated
for the matrix elements $U_{e j}$ of the mixing matrix $U$, taken at
$\theta_{13} = 0$ \cite{Ivanov2} (see also \cite{PDG08}). In the
laboratory frame the $EC$--decay rate is time modulated with a
frequency $\omega_{EC} = \omega_{21}/\gamma$.  Thus, the period
$T_{EC}$ of the time modulation is
\begin{eqnarray}\label{label12}
T_{EC} = \frac{2\pi}{\omega_{EC}}= \frac{2\pi \gamma M_m}{\Delta
m^2_{21}}.
\end{eqnarray}
For the experimental data on the periods of the time modulation
\cite{GSI2}--\cite{GSI5} we get $\Delta m^2_{21} = 2.19\times
10^{-4}\,{\rm eV^2}$ \cite{Ivanov2}.

\subsubsection*{Coherence vs. incoherence in two--body 
electron capture - No interference terms in $EC$--decay rates of
H--like heavy ions \cite{Gal2}}

According to Gal \cite{Gal2}, the two--body K--shell electron capture
$m \to d$ decays, when only the daughter ions are observed, are driven
by the complete set of orthogonal neutrino states
$|\nu_{\alpha}\rangle = \sum_j U^*_{\alpha j} |\nu_j\rangle$ with all
lepton flavours $\alpha = e$, $\mu$ and $\tau$. The amplitude of the
$m \to d + \nu_{\alpha}$ transition is equal to \cite{Gal2}
\begin{eqnarray}\label{label13}
\hspace{-0.15in}A(m \to d\,\nu_{\alpha})(t) = \sum_j U_{\alpha j}U_{e
j}A(m \to d \, \nu_j)(t),
\end{eqnarray}
where the amplitude $A(m \to d\, \nu_j)(t)$ is defined by
Eq.(\ref{label3}).

Since in GSI experiments neutrinos in the $EC$--decays of the H--like
heavy ions are not detected, Gal proposes to define the probability of
the $m \to d$ transition as the incoherent sum of the squared absolute
values of the amplitudes of the $m \to d + \nu_{\alpha}$ transitions
\cite{Gal2}
\begin{eqnarray}\label{label14}
\hspace{-0.15in}&&P(m \to d)(t) = \sum_{\alpha}|A(m \to
d\,\nu_{\alpha})(t)|^2 = \nonumber\\
\hspace{-0.15in}&&= \sum_{\alpha}\sum_i\sum_j U^*_{\alpha i}U^*_{e
i}U_{\alpha j}U_{e j}A^*(m \to d \, \nu_i)(t)\nonumber\\
\hspace{-0.15in}&&\times\,A(m \to d \, \nu_j)(t),
\end{eqnarray}
where the index $\alpha$ runs over $\alpha = e, \mu$ and $\tau$.
Using the orthogonality relation for the matrix elements of the mixing
matrix \cite{PDG08}
\begin{eqnarray}\label{label15}
\sum_{\alpha} U^*_{\alpha
i}U_{\alpha j} = \delta_{ij}
\end{eqnarray}
 one can arrive at the expression
\cite{Gal2}
\begin{eqnarray}\label{label16}
P(m \to d)(t) = \sum_j|U_{e j}|^2|A(m \to d \, \nu_j)(t)|^2,
\end{eqnarray}
which contains no interference term. A similar argument has been
recently given by Yazaki \cite{Yazaki}.

We want to point out here that the use of the complete set of neutrino
wave functions $|\nu_{\alpha}\rangle = \sum_j U^*_{\alpha j}
|\nu_j\rangle$ with all lepton flavours leaves room for the
restoration of the interference terms in the rates of the $m \to d$
transitions.

The amplitude $A(m \to d)(t)$ of the $m \to d$ transition we propose
to define as a coherent superposition of the amplitudes $A(m \to
d\,\nu_{\alpha})(t)$
\begin{eqnarray}\label{label17}
 A(m \to d)(t) = \sum_{\alpha}e^{\,-\,i\varphi_{\alpha}} A(m \to
 d\,\nu_{\alpha})(t),
\end{eqnarray}
where we have introduced arbitrary phases $\varphi_{\alpha}$ for
neutrinos $\nu_{\alpha}$, which are responsible for the restoration of
the interference term.  The amplitudes $A(m \to d\,\nu_{\alpha})(t)$
are determined by Eq.(\ref{label13}). The possibility to describe the
amplitude $A(m\to d)(t)$ by a coherent superposition
Eq.(\ref{label17}) is obvious, since neutrinos $\nu_{\alpha}$ are not
detected.

The rate of the $m \to d$ transition is related to the expression
\cite{Ivanov2}
\begin{eqnarray}\label{label18}
\hspace{-0.3in}&&\lim_{\varepsilon \to
0}\frac{1}{2}\frac{d}{dt}\sum_{M_F}|A(m\to d)(t)|^2 = \nonumber\\
\hspace{-0.3in}&& = 3 M_m |{\cal M}_{\rm GT}|^2|\langle
\psi^{(Z)}_{1s}\rangle|^2 \nonumber\\
\hspace{-0.3in}&&\times\, \Big\{\sum_j\Big|\sum_{\alpha} U_{\alpha
j}U_{ej} e^{\,-\,i\,\varphi_{\alpha}}\Big|^2 2
E_d(\vec{q}\,)E_j(\vec{k}_j)\,2\pi\nonumber\\
\hspace{-0.3in}&&\times \,\delta(\Delta E'_j)\,|\Phi_d(\vec{k}_j
+ \vec{q}\,)|^2 + \sum_{\ell > j}\sqrt{ 2
E_d(\vec{q}\,)E_{\ell}(\vec{k}_{\ell})}\nonumber\\
\hspace{-0.3in}&&\times \sqrt{2
E_d(\vec{q}\,)E_j(\vec{k}_j)}\,[2\pi\,\delta(\Delta E'_{\ell}) +
2\pi\,\delta(\Delta E'_j)]\nonumber\\
\hspace{-0.3in}&&\times\,{\rm Re}\Big[\sum_{\beta}U^*_{\beta
\ell}U^*_{e \ell}\,e^{\,+i\,\varphi_{\beta}}\sum_{\alpha}U_{\alpha
\ell}U_{ej}\,e^{\,-i\,\varphi_{\alpha}} \nonumber\\
\hspace{-0.3in}&&\times\,\Phi^*_d(\vec{k}_{\ell} +
\vec{q}\,)\,\Phi_d(\vec{k}_j + \vec{q}\,)\,e^{\,i\,\omega_{\ell
j}t}\Big]\Big\},
\end{eqnarray}
where indices $\ell$ and $j$ denote the neutrino mass--eigenstates,
the indices $\alpha$ and $\beta$ run over all lepton flavours.  The
frequencies $\omega_{\ell j}$ of the time modulation are defined by
Eq.(\ref{label5}).

As has been remarked in \cite{Ivanov2} and discussed above, the first
term in the r.h.s. of Eq.(\ref{label18}) corresponds to the
decoherent contribution of massive neutrino mass--eigenstates, whereas
the second one is caused by the coherent contribution of the decay
channels $m \to d + \nu_j$. 

The $EC$--decay rates, measured in GSI experiments, take
the form \cite{GSI2}
\begin{eqnarray}\label{label19}
\lambda_{EC}(t) = \lambda_{EC}(1 + a_{EC}\,\cos(\omega_{EC}t +
\phi_{EC})).
\end{eqnarray}
In order to reproduce the correct value of the $EC$--decay constant
$\lambda_{EC}$, calculated in \cite{Ivanov1} in the theory of weak
interactions with massless neutrinos, we impose the following
constraint on the phases of the neutrino wave functions
\begin{eqnarray}\label{label20}
\sum_j\Big|\sum_{\alpha} U_{\alpha j}U_{ej}
e^{\,-\,i\,\varphi_{\alpha}}\Big|^2 = 1.
\end{eqnarray}
Setting the mixing angels $\theta_{13} = 0$ and $\theta_{23} = \pi/4$
and using the definition of the mixing matrix $U$ \cite{PDG08}, the
condition Eq.(\ref{label20}) can be transcribed into the form
\begin{eqnarray}\label{label21}
  \hspace{-0.3in}&&\sum_j\Big|\sum_{\alpha} U_{\alpha j}U_{ej}
  e^{\,-\,i\,\varphi_{\alpha}}\Big|^2 =  1\nonumber\\
  \hspace{-0.3in}&& - \frac{1}{2}\,
\sin^22\theta_{12}\cos(\varphi_{\mu e} - \varphi_{\tau e}) + 
\frac{1}{\sqrt{2}}\,\sin 2\theta_{12}\nonumber\\
  \hspace{-0.3in}&&\times\,(\cos \varphi_{\tau e} - \cos \varphi_{\mu
e}) \cos 2\theta_{12} \, = 1,
\end{eqnarray}
where $\varphi_{\alpha e} = \varphi_{\alpha} - \varphi_e$ for $\alpha
= \mu, \tau$.

The interference term is
\begin{eqnarray}\label{label22}
\hspace{-0.3in}&&\sum_{\ell > j} 2{\rm Re}\Big[\sum_{\beta \,
\alpha}U^*_{\beta \ell}U^*_{e \ell}
e^{i(\varphi_{\beta}-\varphi_{\alpha})} U_{\alpha \ell}U_{ej}e^{
i\omega_{\ell j}t}\Big] =\nonumber\\
\hspace{-0.3in}&&= \frac{1}{\sqrt{2}}\,\sin2\theta_{12}\,(\sin\varphi_{\mu e} -
\sin\varphi_{\tau e})\,\sin(\omega_{21} t).
\end{eqnarray}
where we have used Eq.(\ref{label21}).  

Identifying the rate of the $m \to d$ transition with the $EC$--decay
rate Eq.(\ref{label19}) we get
\begin{eqnarray}\label{label23}
  \omega_{EC} &=& \frac{\omega_{21}}{\gamma} = \frac{\Delta
  m^2_{21}}{2 \gamma M_m},\nonumber\\ \phi_{EC} &=&
  -\,\frac{\pi}{2},\nonumber\\ a_{EC} &=&
  \frac{1}{\sqrt{2}}\,\sin2\theta_{12}\,(\sin \varphi_{\mu e} - \sin
  \varphi_{\tau e}).
\end{eqnarray}
Thus, we predict that the phase and amplitude of the time modulated
term of the $EC$--decay rate of the H--like heavy ions should be
universal and equal to $\phi_{EC} = - \pi/2$ and $a_{EC} =
\frac{1}{\sqrt{2}}\,\sin2\theta_{12}\,(\sin \varphi_{\mu e} - \sin
\varphi_{\tau e})$ (see Eq.(\ref{label23})), respectively.  

For the experimental value of the mixing angle $\theta_{12} = 34^0$
\cite{PDG08} and the modulation amplitude $a_{EC} = 0.21$
\cite{GSI2}--\cite{GSI5} we get the following system of equations
\begin{eqnarray}\label{label24}
\hspace{-0.3in}&&\Big\{\begin{array}{r@{ = }l} \sin \varphi_{\mu e} - \sin
\varphi_{\tau e}\, &\, 0.32 \\ 1.75\,\cos(\varphi_{\mu e} - \varphi_{\tau
e}) + \cos \varphi_{\mu e} - \cos \varphi_{\tau e}\, &\, 0
\end{array},\nonumber\\
\hspace{-0.3in}&&
\end{eqnarray}
defining the phase differences $\varphi_{\mu e}$ and $\varphi_{\tau
e}$.  The second equation we have obtained from the condition
Eq.(\ref{label21}). One of the solutions of the system
Eq.(\ref{label24}) is $\varphi_{\mu e} \simeq 1.75\,{\rm rad}$ and
$\varphi_{\tau e} \simeq 0.73\,{\rm rad}$.

For the understanding of the physical origin of the phases
$\varphi_{\alpha}$ one can analyse, for example, the mixing matrix for
neutrinos $\nu_{\alpha}$ with definite lepton charges $\alpha = e,\mu$
and $\tau$ as coherent superpositions of massive neutrino
mass--eigenstates. The wave functions of neutrinos $\nu_{\alpha}$,
which we use for the calculation of the amplitudes of the $EC$--decay
rates, can be written as $|\nu_{\alpha}\rangle =
\sum_j\tilde{U}^*_{\alpha j}|\nu_j\rangle$, where the mixing matrix
$\tilde{U}^*$ is defined by
\begin{eqnarray}\label{label25}
\hspace{-0.3in}&&\tilde{U}^* =  \left(\begin{array}{cccc} e^{\,i\varphi_e}  &
        0 & 0 \\ 0 & e^{\,i\varphi_{\mu}} & 0\\ 0 & 0 &
        e^{\,i\varphi_{\tau}}\\
\end{array}\right) U^* = \nonumber\\
\hspace{-0.3in}&& = \left(\begin{array}{cccc} e^{\,i\varphi_e} & 0 & 0
        \\ 0 & e^{\,i\varphi_{\mu}} & 0\\ 0 & 0 &
        e^{\,i\varphi_{\tau}}\\
\end{array}\right)\left(\begin{array}{cccc}  c_{12} &
        s_{12}& 0 \\ {\displaystyle - \,\frac{s_{12}}{\sqrt{2}}}
        &{\displaystyle +\,\frac{c_{12}}{\sqrt{2}}} & {\displaystyle
        \frac{1}{\sqrt{2}}}\\ {\displaystyle
        +\,\frac{s_{12}}{\sqrt{2}}} &{\displaystyle -
        \frac{c_{12}}{\sqrt{2}}}&{\displaystyle \frac{1}{\sqrt{2}}}\\
    \end{array}\right).\nonumber\\
\hspace{-0.3in}&&
\end{eqnarray}
Here $U^*$ is the standard mixing matrix, calculated for the mixing
angles $\theta_{13} = 0$ and $\theta_{23} = \pi/4$ \cite{PDG08}.  In
such a representation of the mixing matrix, the phase differences
$\varphi_{\mu e} = \varphi_{\mu} - \varphi_e$ and $\varphi_{\tau e} =
\varphi_{\tau} - \varphi_e$ may have the meaning of Majorana phases,
caused by CP--violation \cite{PDG08}.

Since due to the condition Eq.(\ref{label21}) the limit of equal
phases $\varphi_e = \varphi_{\mu} = \varphi_{\tau} = \varphi$,
including $\varphi = 0$, of wave functions of neutrinos with lepton
flavours $\alpha = e,\mu$ and $\tau$ corresponds to the massless limit
of massive neutrino mass--eigenstates $m_j \to 0$, being equivalent to
the vanishing of the mixing angle $\theta_{12} \to 0$, the
interference term vanishes for $\varphi_e = \varphi_{\mu} =
\varphi_{\tau} = \varphi$ and, of course, for $\varphi = 0$ as a
partial case.

\subsubsection*{Final state interference in $EC$--decay rates 
is due to the neutrino magnetic moment \cite{Gal2}}

The ``GSI oscillations'' cannot be induced by the magnetic moments of
neutrinos, since in this case the experimental data on the $EC$ and
$\beta^+$ decays of the H--like ions should show a time modulation
with equal periods. This contradicts \cite{GSI2}--\cite{GSI5}.

\subsubsection*{Conclusive discussion}

We have shown that the appearance of the interference terms in the
$EC$--decay rates of the H--like heavy ions with periods, proportional
to the mass of the H--like mother heavy ion $T_{EC} \sim M_m$ but not
the $Q$--value of the $EC$--decay, is due to overlap of massive
neutrino mass--eigenstate energies and of the wave functions of the
daughter ions in two--body decay channels $m \to d + \nu_1$ and $m \to
d + \nu_2$, caused by the energy and momentum uncertainties,
introduced by the time differential detection of the daughter ions in
GSI experiments.

We have shown that the idea that neutrinos with all lepton flavours
contribute to the $EC$--decays of the H--like heavy ions, which has
been used by Gal to show the non--existence of interference terms, can
be adopted for the derivation of the $EC$--decay rates with
interference terms.

We have pointed out that magnetic moments of massive neutrino
mass--eigenstates cannot be responsible for the time modulation of the
$EC$--decay rates of the H--like have ions. The most important
objective is that such a time modulation should be universal for $EC$
and $\beta^+$ decay rates of all H--like heavy ions. This contradicts
the experimental data \cite{GSI2}--\cite{GSI5}.

This research was partly supported by the DFG cluster
of excellence "Origin and Structure of the Universe" of the Technische
Universit\"at M\"unchen and the Austrian ``Fonds zur F\"orderung
der Wissenschaftlichen Forschung'' (FWF) under contract P19487-N16.

\end{document}